\documentclass[a4paper,11pt,twocolumn,twoside]{article}
\usepackage[dvips]{graphicx}
\usepackage{sepln_en}
\usepackage{fullname}
\usepackage[utf8]{inputenc}
\usepackage{tabularx}
\usepackage{adjustbox}
\usepackage{subcaption}
\usepackage{booktabs}

\usepackage{color}

\usepackage{xcolor}
\definecolor{bildu}{HTML}{b0d136}
\definecolor{pnv}{HTML}{439d39}
\definecolor{mg}{HTML}{2429a6}
\definecolor{bng}{HTML}{52ffdc}
\definecolor{erc}{HTML}{ffd21f}
\definecolor{jxc}{HTML}{52ffdc}
\definecolor{cup}{HTML}{f8ff1f}
\definecolor{psoe}{HTML}{e30613}
\definecolor{pp}{HTML}{0bb2ff}
\definecolor{up}{HTML}{cd37c8}
\definecolor{cs}{HTML}{fa5000}

\input epsf

\title{Political Leaning Inference through Plurinational Scenarios}

\author {\textbf{Joseba Fernandez de Landa,$^1$} \textbf{Rodrigo Agerri$^1$}\\
$^1$HiTZ Center - Ixa, University of the Basque Country UPV/EHU\\
joseba.fernandezdelanda@ehu.eus\\}

\seplnkey{Political analysis, Computational Social Science, Social Media}

\seplnabstract{
Social media users express their political preferences via interaction with other users, by spontaneous declarations or by participation in communities within the network. This makes a social network such as Twitter a valuable data source to study computational science approaches to political learning inference. In this work we focus on three diverse regions in Spain (Basque Country, Catalonia and Galicia) to explore various methods for multi-party categorization, required to analyze evolving and complex political landscapes, and compare it with binary left-right approaches. We use a two-step method involving unsupervised user representations obtained from the retweets and their subsequent use for political leaning detection. Comprehensive experimentation on a newly collected and curated dataset comprising labeled users and their interactions demonstrate the effectiveness of using Relational Embeddings as representation method for political ideology detection in both binary and multi-party frameworks, even with limited training data. Finally, data visualization illustrates the ability of the Relational Embeddings to capture intricate intra-group and inter-group political affinities.}

\firstpageno{1}

\begin{document}

\label{firstpage} 
\maketitle

\section{Introduction}

Social media users spontaneously engage in political activities, sharing their generated content and interacting with other users from similar or different ideology. As such, social media has become a valuable data source for researchers to study political communities \cite{conover2011political,Barber2015BirdsOT,garimella2017long}, understand partisanship \cite{Leong2020ConservativeAL}, polarization \cite{Morales2021NoEI}, and other related political phenomena.

Many computational science studies of political behaviour in social media have assumed that it is feasible to reliably infer political leanings by automatic methods. However, most of the research on inferring political leaning has predominantly focused on binary orientations such as left-right \cite{conover2011political,Barber2015UnderstandingTP,Barber2015BirdsOT,garimella2017long,conover2011predicting}, liberal-conservative \cite{barbera2015tweeting,PreotiucPietro2017BeyondBL} and even democrat-republican \cite{Pennacchiotti2011DemocratsRA,Hua2020TowardsMA,timme20}. In fact, the limited number of works that considered a wider leaning spectrum are limited to a specific region \cite{Boutet2012WhatsIY,Makazhanov2013PredictingPP,Rashed2021EmbeddingsBasedCF}. This binary and region-specific outlook represents the political landscape as a uniform and static entity. In reality, every social context has its unique political landscape, characterized by more than just two ideological choices, which evolve in response to societal requirements.

In other words, political leaning inference remains a challenging problem, specially for social media users which are not actively engaged in political activities or that very infrequently express their political preferences. Furthermore, traditional studies on binary orientations are not able to capture the wide spectrum of political ideology and nuances in traditional two-party political systems which are evolving into scenarios where new political actors are emerging \cite{lisi2018party} as a response to new ideological conflicts \cite{ford2020changing}, socio-economic consequences \cite{kotroyannos2018south,morlino2017impact} or by suggesting innovative political proposals and novel approaches \cite{endbipartspain}. 

Furthermore, transnational integration is leading to the emergence of plurinational states, where the presence of a singular national sentiment is not readily apparent \cite{KeatingMichael}, and diverse political leanings representing distinct national sentiments are arising \cite{MCGANN201948}. Consequently, each political region develops its own political parties tailored to suit specific socio-political circumstances with the aim of obtaining support and sympathy among the population. Hence, characterizing political leaning in terms of proximity to a specific political party would offer a more accurate representation of political nuances in contrast to oversimplifying frameworks such as left-right or conservative-liberal. Moreover, a dynamic approach is essential for tailoring political leaning inference to specific times and regions, specially on complex political scenarios characterized by numerous and evolving political options. We believe that this paradigm could enrich opinion polls, political polarization studies, stance detection and also achieve more accurate social and political research.

To effectively deal with the complexities of real-world politics, our investigation delves into the multi-party framework outlined by \namecite{FdLPolLeaIn}, where the main challenge is to establish effective and robust automatic user representation methods able to capture complex socio-political information for a number of political leaning inference tasks, including left-right or multi-party political leaning, among others. In this work users are represented via their retweet interactions \cite{conover2011predicting,magdy16,darwish2020unsupervised,Stefanov2020PredictingTT,fernandez2022relational}. 

More specifically, a number of unsupervised techniques to generate user representations are explored: Relational Embeddings \cite{fernandez2022relational}, ForceAtlas2 \cite{jacomy2014forceatlas2}, DeepWalk \cite{deepwalk} and Node2vec \cite{Grover2016node2vecSF}. In this paper these language independent user representation techniques are evaluated for their effectiveness in inferring binary and multi-party political leanings within three different politically complex regions in Spain, namely, Basque Country, Catalonia and Galicia.

To this end, this paper makes the following contributions on political leaning inference: 
(i) a novel publicly available dataset containing labeled users by political party and left-right orientation alongside their retweets from the regions of Basque Country, Catalonia and Galicia;
(ii) comprehensive experimentation with multi-party (7 political parties) and binary (left-right) frameworks for the three aforementioned political regions, showing that Relational Embeddings \cite{fernandez2022relational} outperform other user representation methods, especially in few-shot evaluation settings;
(iii) an error analysis and data visualization illustrate the potential of our method to infer political ideology, even in dynamic and politically complex scenarios;
(iv) data and code will be available upon publication.

\section{Related work} 

Social media is presented as a source to extract data that can then be used to represent users. One of the most common approaches to transform social media interactions into these users representations is based on the force-directed algorithm \cite{Fruchterman1991GraphDB}. The force-directed algorithm has been used to create unsupervised user representations and subsequently perform stance detection on Twitter users \cite{darwish2020unsupervised}. ForceAtlas2 \cite{jacomy2014forceatlas2}, a forced-directed algorithm, has also been widely used to represent latent user structures. For instance, to study political misinformation in the 2018 presidential election in Brazil, based on opposed hashtags \cite{soares2021hashtag}, or to identify the roles that users play in political conversations during polarized online discussions \cite{recuero2019using}. Similar techniques were used to map Persian Twitter during Iran's 2017 presidential election by investigating the network structures generated by the users and their sharing practices \cite{kermani2021mapping}. Furthermore, the ForceAtlas2 method has also been used to represent latent user structures through interactions to analyze affinities towards independence movements \cite{Zubiaga2019PoliticalHI}, young Basque communities \cite{fernandez2019large} or left-right alignments \cite{conover2011predicting}. These methods convert large interaction matrices into two-dimensional features, significantly reducing sparsity and memory usage, albeit at the cost of losing some information.

Neural approaches for unsupervised user or node representation based on connections include DeepWalk \cite{deepwalk} and node2vec \cite{Grover2016node2vecSF} as the most popular and effective methods \cite{socialphysics2022,9565320}, also to represent users via social media interactions \cite{basqueyoungtwitter,fernandez2022relational,QMUL-SDS}. The node-representation features are generated from unlabeled data, representing nodes as low-dimensional features, and they try to predict a group of neighboring users that emerge from Random Walks, based on the input of a given user \cite{mikolov2013efficient}. More recently, Relational Embeddings have been proposed as an alternative technique to generate user representations based on real user-to-user interaction pairs \cite{fernandez2022relational}.

\section{Datasets}\label{sec:datacol}

We build a dataset to classify multiparty political ideology in three different regions of Spain, namely, Basque Country, Catalonia and Galicia. First, we select the most relevant political parties for each of them. Subsequently, we extract data from Twitter, manually labeling users and collecting the interaction data required to build the user representations.

\subsection{Political Regions}\label{sec:tar_sel}

Given our interest in analyzing complex political contexts, we have chosen the Kingdom of Spain on 2020 summer as our case of study. Spain is a complex political context, characterized not only by its status as a plurinational country \cite{KeatingMichael} but also because of the emergence of new political actors in the last years \cite{endbipartspain}. Thus, new political actors (UP, Cs, and Vox) burst into Spanish political scenario ruled by traditional forces (PSOE and PP), suggesting updated political proposals and novel approaches \cite{endbipartspain}. Moreover, our study focuses on the Basque Country, Catalonia and Galicia, which are considered stateless nations within the multinational state of Spain \cite{KeatingMichael}. These regions house a greater number of political parties than other areas, each with its own regional parliaments and distinct nationalist orientations that drive a wide range of political choices. For each of the selected regions we have picked the political parties that have (or potentially may have) political representation on the regional parliaments. In order to compare multi-party to a binary framework, we also annotate each of the political parties with left-right labels.

\textbf{Basque Country (EUS):} \textit{Basque Nationalist Party} (Partido Nacionalista Vasco - PNV \textcolor{pnv}{$\bullet$}) Basque nationalist and Christian-democratic political party; \textit{Unite} (EH Bildu \textcolor{bildu}{$\bullet$}) left-wing Basque pro-independence coalition; \textit{Socialist Party} (Partido Socialista Obrero Español - PSOE \textcolor{psoe}{$\bullet$}) social-democratic Spanish political party; \textit{Together We Can} (Unidos Podemos - UP \textcolor{up}{$\bullet$}) democratic socialist Spanish electoral alliance; \textit{People's Party} (Partido Popular - PP \textcolor{pp}{$\bullet$}) conservative and Christian-democratic Spanish political party; \textit{Citizens} (Ciudadanos - Cs \textcolor{cs}{$\bullet$}) liberal Spanish political party; \textit{Vox} (Vox $\bullet$) conservative Spanish political party.

\textbf{Galicia (GAL):} \textit{Galician Nationalist Bloc} (Bloque Nacionalista Galego - BNG \textcolor{bng}{$\bullet$}) left-wing Galician nationalist coalition; \textit{Galicianist Tide} (Marea Galeguista - MG \textcolor{mg}{$\bullet$}) left-wing Galician electoral alliance. Despite some punctual differences for PSOE, UP, PP, Cs and Vox are the same as the Basque Country above, whereas representatives and party accounts collected are specific for this region.

\textbf{Catalonia (CAT):} \textit{Republican Left of Catalonia} (Esquerra Republicana de Catalunya - ERC \textcolor{erc}{$\bullet$}) social-democratic Catalan pro-independence political party; \textit{Together for Catalonia} (Junts per Catalunya - JxC \textcolor{jxc}{$\bullet$}) progressives Catalan pro-independence political party; \textit{Popular Unity Candidacy} (Candidatura d'Unitat Popular - CUP \textcolor{cup}{$\bullet$}) left-wing Catalan pro-independence political party. PSOE, UP, PP and Cs are the same as for Basque Country and Galicia, although the collected representatives and party accounts are specific for Catalonia. Note that at the time of data collection the Vox party did not have yet representation.

\textbf{Binary framework:} In addition to the political party identification, we have incorporated a binary categorization to determine each party's left-right alignment. This was done to facilitate a comparison between the proposed multi-party framework, grounded in political parties, and a more straightforward binary framework. To categorize each political party, we utilized data from opinion polls \cite{cis_es,cis_gal,cis_eus}, which gauge public perceptions of the left-right orientations of these parties.

\subsection{Data collection strategy}
\label{sec:colest}

Our data collection strategy consists of two steps, each applied to the selected regions. The initial step involves labeling users for supervised learning, while the second step focuses on gathering interaction data that will be use as input data for user representations.

\noindent (1) \textbf{Manual labeling:}
Our starting point relies on a user's seed list that consists of users related to the selected political parties of each region. Thus, we are selecting a sample of users in order to collect data, following the same technique as done in other works \cite{Makazhanov2013PredictingPP,Barber2015BirdsOT,garimella2017long,Hua2020TowardsMA}. The selected users are related to political parties of each region, such as the political organizations, elected members, candidates or even political militants. The identified user's are labeled by its political party, forming our labeled sample. Apart for the party level categorization for the multi-party framework, we add a binary categorization as left (L) or right (R) leaning for the binary framework (see Table \ref{tab:data_labels}). In the following steps, this user list is going to be used to extract the interactions to build user features for the labeled users. 

\begin{table}[h!]\tiny
\centering
\begin{tabular}{@{}rr|rr|rr@{}}
\toprule
\multicolumn{2}{c}{EUS}  & \multicolumn{2}{c}{GAL}  & \multicolumn{2}{c}{CAT}   \\ \midrule
party & n            &  party & n           &  party & n  \\ \midrule                   
PNV (R)     \textcolor{pnv}{$\bullet$}      & 146                
& BNG (L)   \textcolor{bng}{$\bullet$}      & 39                 
& ERC (L)   \textcolor{erc}{$\bullet$}      & 18 \\
Bildu (L)   \textcolor{bildu}{$\bullet$}    & 134              
& MG (L)    \textcolor{mg}{$\bullet$}       & 7                   
& JxC (R)   \textcolor{jxc}{$\bullet$}      & 18 \\
UP (L)      \textcolor{up}{$\bullet$}       & 177                 
& UP (L)    \textcolor{up}{$\bullet$}       & 48 
& UP (L)    \textcolor{up}{$\bullet$}       & 16 \\
PSOE (L)    \textcolor{psoe}{$\bullet$}     & 157               
& PSOE (L)  \textcolor{psoe}{$\bullet$}     & 35                
& PSOE (L)  \textcolor{psoe}{$\bullet$}     & 18 \\
PP (R)      \textcolor{pp}{$\bullet$}       & 132                 
& PP (R)    \textcolor{pp}{$\bullet$}       & 45                  
& PP (R)    \textcolor{pp}{$\bullet$}       & 14 \\
Cs (R)      \textcolor{cs}{$\bullet$}       & 40                  
& Cs (R)    \textcolor{cs}{$\bullet$}       & 12                  
& Cs (R)    \textcolor{cs}{$\bullet$}       & 14 \\
Vox (R)     $\bullet$                       & 8                  
& Vox (R)   $\bullet$                       & 7                  
& CUP (L)   \textcolor{cup}{$\bullet$}      & 11 \\ \midrule
TOTAL:&794               &TOTAL:&193                &TOTAL:&109\\ \bottomrule
\end{tabular}
\caption{\scriptsize Labeled users for each region. Columns correspond to political party and number of users (n) per region. The labels between parenthesis correspond to left (L) or right (R) leaning for the binary framework.}
\label{tab:data_labels}
\end{table}

\noindent (2) \textbf{Twitter data retrieval:} 
For every labeled user we first retrieve a history of their retweet interactions. The purpose of this initial retrieval is not to gather the retweets themselves but to identify all users who have interacted with the labeled users through retweets. Afterwards, we gathered all retweets accessible from the timelines the users, extracting a substantial number of interactions involving the sharing of content among users. The data obtained from the timeline extraction ensures that we have enough quantity to later represent the labeled users as well as the users interacting with them \cite{fernandez2022relational}. Table \ref{tab:data_final} shows the number of retweets retrieved from the labeled and interacting users during the summer of 2020.

\begin{table}[h!] \scriptsize 
\centering
\begin{tabular}{@{}lrrr@{}}
\toprule
                     & EUS       & GAL       & CAT   \\ \midrule
Labeled users        & 794       & 193       & 109   \\ \midrule
Interacting users    & 155K     & 50K     & 144K    \\ 
Retweets             & 58M     & 13M      & 41M  \\ \bottomrule
\end{tabular}
\caption{Final dataset composition for each region.}
\label{tab:data_final}
\end{table}

\section{Method}

The general idea consists of building user representation features able to capture sociopolitical information leveraging social media data, emulating the same methodology presented in \namecite{FdLPolLeaIn} and extending it to complex scenarios. These representations will be exclusively built from retweets, without any textual data, as research has shown the effectiveness of retweet-based interactions in user classification tasks \cite{conover2011predicting,magdy16,darwish2020unsupervised,Stefanov2020PredictingTT,fernandez2022relational,cignarella2020sardistance,vaxxstance2021}. In order to compress that information from sparse interaction matrices into dense user representations, 
the selected unsupervised training strategies act like feature extraction methods, representing each user in a dense vector space. These methods are used to effectively transform a large number of content sharing actions into meaningful sociopolitical user representations. Those features will be subsequently employed alongside different classification algorithms to evaluate their performance for inferring the political leaning of the labeled users.

\subsection{User Representations}

We have chosen four different unsupervised approaches to extract user features from interactions. Instead of leveraging huge sparse adjacency matrices, user interactions are brought into a low dimensional vector space, based on approximation-repulsion \cite{Fruchterman1991GraphDB} or context embeddings \cite{mikolov2013efficient}. We employ the following methods for comparison:

\noindent - \textbf{ForceAtlas2} \cite{jacomy2014forceatlas2} (FA2) operates repulsing unconnected nodes and attracting connected ones in order to generate a 2 dimensional graph.

\noindent - \textbf{DeepWalk} \cite{deepwalk} (DW) simulates uniform random walks among connected instances from a network to learn feature representations. 

\noindent - \textbf{Node2vec} \cite{Grover2016node2vecSF} (N2V) algorithm is similar to DeepWalk, but they add control parameters to control the structure of the generated graph. 

\noindent - \textbf{Relational Embeddings} \cite{fernandez2022relational} (RE) are based on a single hidden-layer neural network trying to predict who retweeted whom for all the gathered interaction pairs. Instead of generating random walks among nearest neighbours, in this method the interactions are based on the relations between two users.

\subsection{Methodology}

Separate user representations are generated with the four methods listed above, exclusively utilizing retweet data and excluding any labels. Independent models are trained for EUS, GAL and CAT regions without data mixing across them, with the intention of keeping the idiosyncrasies of each region unaltered. All the available retweet interactions per region are employed without any filtering for every method. Feature dimensions are set on 20 dims for DeepWalk, node2vec and Relational Embeddings, with the intention of generating low dimensional but meaningful features \cite{darwish2020unsupervised,Stefanov2020PredictingTT,fernandez2022relational}. The parameter settings used for node2vec and DeepWalk are the default values typically used by these algorithms: walks\_per\_node = 10, walk\_length = 80, window or context\_size = 10, and the optimization is run for a single epoch \cite{deepwalk,Grover2016node2vecSF,fernandez2022relational}. Specifically for node2vec, we set p=1 and q=0.5 in order to enhance network community related information \cite{Grover2016node2vecSF}. Default parameters were set for ForceAtlas2.

\section{Experimental setup} 

In order to identify the political leaning of the labeled users among EUS, GAL and CAT regions, we represent each user with the vectors arisen from the different user representations. To evaluate the performance of these generated user representations, we will conduct two separate sets of experiments with the aim of inferring political leaning. The first experiment will compare the different user representation methods when inferring both the left-right orientation (binary) or political leaning (multi-party) of each user within a strongly supervised scenario. The second experiment will involve a more challenging weakly supervised scenario, where a classifier is provided with very limited training data.

\subsection{Strongly supervised scenario} 
\label{sec:str_ss}

The main goal of this experiment is to compare the performance of various user representation methods when applying the same classifiers. Furthermore, we aim to compare our approach, which defines political leaning within a dynamic multi-party framework, with the conventional approach of treating it as a binary categorization. Thus, we experiment with different user representation methods while inferring binary or multi-class political leanings: (i) \textit{Binary framework}, where only two classes are used to define users' political orientation as left or right. These same categories are used across regions, making it a generalizable and uniform approach. (ii) \textit{Multi-party framework}, where users' political leaning is defined as the closest political party. Political parties vary by region, with seven parties in each region, as specified in Section \ref{sec:tar_sel}.

For each region and framework, we conduct experiments using a leave-one-out (LOO) cross-validation approach. This means that one user is held out for testing while all the remaining users are utilized for training. Therefore, a model is trained and tested individually for every user in the dataset, a feasible task due to the low dimensionality of the representations. The user representations obtained from each of the methods are used to train six different classification algorithms for each region and framework: Logistic Regression (LogReg), Random Forest (RF), Naive Bayes (NB) and linear, polynomial and RBF-kernel Support Vector Machines (SVM). We use the Scikit-learn implementation \cite{scikit-learn} with default configurations.

\begin{table*}[ht!]\tiny%
\centering
\begin{tabular}{@{}l|llll|llll|llll|llll@{}}
\toprule
\multicolumn{1}{c}{}
& \multicolumn{4}{c}{EUS}  & \multicolumn{4}{c}{GAL}  & \multicolumn{4}{c}{CAT} & \multicolumn{4}{c}{average} \\ \midrule
&N2V&DW&FA2&RE  &N2V&DW&FA2&RE &N2V&DW&FA2&RE &N2V&DW&FA2&RE \\ \midrule
LogReg   
& 83.7  & 87.2  & 76.3  & \textbf{96.0} 
& 87.1  & 91.4  & 97.7  & \textbf{98.2} 
& 97.2  & \textbf{\underline{99.1}}  & 69.5  & 98.1 
& 89.3	& 92.6	& 81.2	& \textbf{97.4} 
\\
RF    
& 82.7  & 87.3  & 83.0  & \textbf{\underline{96.5}} 
& 91.9  & 97.1  & \textbf{\underline{98.8}}  & 98.2 
& 96.2  & \textbf{98.1}  & 89.7  & 96.2 
& 90.3	& 94.2	& 90.5	& \textbf{97.0} 
\\
NB        
& 49.9  & 50.3  & 74.5  & \textbf{93.0} 
& 59.0  & 60.1  & 97.6  & \textbf{98.2} 
& 69.3  & 63.6  & 74.6  & \textbf{98.1} 
& 59.4	& 58.0	& 82.2	& \textbf{96.4} 
\\
SVM - lin.    
& 80.5  & 85.5  & 77.2  & \textbf{95.4} 
& 85.0  & 89.4  & \textbf{98.3}  & 98.2
& 96.2  & \textbf{98.1}  & 73.7  & 97.2 
& 87.2	& 91.0	& 83.1	& \textbf{96.9}
\\
SVM - pol.     
& 57.4  & 59.7  & 82.9  & \textbf{93.9} 
& 41.7  & 43.3  & \textbf{\underline{98.8}}  & 96.4 
& 88.0  & 87.1  & 74.6  & \textbf{98.1} 
& 62.4	& 63.4	& 85.4	& \textbf{96.1} 
\\
SVM - rbf       
& 69.5  & 73.9  & 83.5  & \textbf{95.1} 
& 64.4  & 76.1  & \textbf{98.3}  & 98.2 
& 84.0  & 86.8  & 74.6  & \textbf{98.1}
& 72.6	& 78.9	& 85.5	& \textbf{97.1} 
\\ \midrule
average
& 70.6	& 74.0	& 79.6	& \textbf{95.0} 
& 71.5	& 76.2	& \textbf{98.3}	& 97.9 
& 88.5	& 88.8	& 76.1	& \textbf{97.6} 
& 76.9	& 79.7	& 84.6	& \textbf{96.8} 
\\ \bottomrule
\end{tabular}
\caption{\scriptsize BINARY FRAMEWORK (left-right). F1 macro score results for strongly supervised scenario (LOO CV) on EUS, GAL and CAT datasets. Values in \textbf{bold} represent best results for each classifier, while \underline{underlined} values represent best overall results for each dataset.}
\label{tab:f1_strong_bi}
\end{table*}

\begin{table*}[ht!]\tiny%
\centering
\begin{tabular}{@{}l|llll|llll|llll|llll@{}}
\toprule
\multicolumn{1}{c}{}
& \multicolumn{4}{c}{EUS}  & \multicolumn{4}{c}{GAL}  & \multicolumn{4}{c}{CAT} & \multicolumn{4}{c}{average} \\ \midrule
&N2V&DW&FA2&RE  &N2V&DW&FA2&RE &N2V&DW&FA2&RE &N2V&DW&FA2&RE \\ \midrule
LogReg   
& 63.7  & 66.1  & 33.3  & \textbf{\underline{94.0}}
& 42.2  & 56.4  & 46.6  & \textbf{92.6}
& 86.7  & 88.0  & 33.6  & \textbf{95.1}
& 64.2	& 70.2	& 37.8	& \textbf{93.9}
\\
RF    
& 59.8  & 70.1  & 50.6  & \textbf{92.8}
& 62.2  & 71.3  & 63.2  & \textbf{90.1}
& 77.2  & 80.3  & 47.8  & \textbf{91.6} 
& 66.4	& 73.9	& 53.9	& \textbf{91.5}
\\
NB        
& 38.5  & 41.5  & 42.6  & \textbf{86.7}
& 46.4  & 46.8  & 50.3  & \textbf{87.9}
& 39.5  & 53.9  & 63.1  & \textbf{91.8} 
& 41.5	& 47.4	& 52.0	& \textbf{88.8}
\\
SVM - lin.    
& 61.3  & 64.6  & 33.8  & \textbf{93.1}
& 38.6  & 53.9  & 47.0  & \textbf{\underline{92.8}}
& 82.5  & 86.4  & 21.0  & \textbf{\underline{96.2}} 
& 60.8	& 68.3	& 33.9	& \textbf{94.0}
\\
SVM - pol.     
& 36.1  & 37.1  & 42.2  & \textbf{87.7}
& 06.2  & 08.4  & 47.3  & \textbf{89.9}
& 59.4  & 71.1  & 07.1  & \textbf{93.7} 
& 33.9	& 38.9	& 32.2	& \textbf{90.4}
\\
SVM - rbf       
& 39.6  & 42.1  & 48.6  & \textbf{92.3}
& 20.5  & 28.1  & 47.4  & \textbf{91.7}
& 52.2  & 59.2  & 18.7  & \textbf{94.2} 
& 37.4	& 43.1	& 38.2	& \textbf{92.7}
\\ \midrule
average
& 49.8	& 53.6	& 41.9	& \textbf{91.1}	
& 36.0	& 44.2	& 50.3	& \textbf{90.8}	
& 66.3	& 73.2	& 31.9	& \textbf{93.8}
& 50.7	& 57.0	& 41.3	& \textbf{91.9}
\\ \bottomrule
\end{tabular}
\caption{\scriptsize MULTI-PARTY FRAMEWORK (7 political parties). F1 macro score results for strongly supervised scenario (LOO CV) on EUS, GAL and CAT datasets. Values in \textbf{bold} represent best results for each classifier, while \underline{underlined} values represent best overall results for each dataset.}
\label{tab:f1_strong_mul}
\end{table*}

\subsection{Weakly supervised scenario} 
\label{sec:wk_ss} 

Next we experiment in a more challenging, weakly supervised scenario, where the classifiers are provided with limited training data for each region in two different settings: (i) \textit{One-shot}, where only one item per class is selected for training. The selected item for each class is manually selected, being the item representing each of the political parties. (ii) \textit{Three-shot}, a few-shot setting where three items per class are selected for training. In this occasion, for each class we will select a single user corresponding to the political party, as well as two users representing the most referential candidates.

The remainder of the users are left for the test set. In this scenario, the inference will be conducted at the political party level within the multi-party framework. For this setting we only use the RE user representations and SVM-linear classifier, which achieved the best results in the multi-party framework for the strongly supervised scenario (94.0 F1 macro score, Table \ref{tab:f1_strong_mul}). Additionally, we provide the results obtained both with and without employing dimensionality reduction techniques. The dimensions have been reduced to 2, in accordance with prior research \cite{darwish2020unsupervised,Stefanov2020PredictingTT}, while the remaining hyperparameters are set to their default values. Three different dimensionality reduction techniques are used for this purpose, PCA, t-SNE \cite{van2008visualizing} and UMAP \cite{McInnes2018UMAPUM}.

\begin{table*}[ht!]\scriptsize
\centering
\begin{tabular}{@{}l||l|rrr||l|rrr||l|rrr@{}}
\toprule
\multicolumn{1}{c}{}
& \multicolumn{4}{c}{EUS}  & \multicolumn{4}{c}{GAL}  & \multicolumn{4}{c}{CAT} \\ \midrule
Dim. Red. &LOO&3-shot&1-shot&\textit{avg}  &LOO&3-shot&1-shot&\textit{avg} &LOO&3-shot&1-shot&\textit{avg} \\ \midrule
none   
&\textbf{93.1} &         *76.5    &       55.4     &       66.0 
&\textbf{92.8} &         *81.4    &*\textbf{88.5}  &       85.0 
&\textbf{96.2} &*\textbf{94.2}    &       *89.2    &       91.7 
\\
UMAP  2d    
&        89.5  &*\textbf{90.0}    &*\textbf{81.6}  &       \textbf{85.8} 
&        80.2  &         *83.5    &       *82.4    &       83.0 
&        95.4  &*\textbf{94.2}    &*\textbf{95.1}  &       \textbf{94.7} 
\\
t-SNE 2d        
&        90.5  &         *77.7    &       *72.9    &       75.3 
&        85.1  &*\textbf{85.5}    &       *86.8    &       \textbf{86.2} 
&        94.3  &         *93.3    &*\textbf{95.1}  &       94.2 
\\
PCA   2d    
&        49.9   &         28.2     &       26.6     &       27.4 
&        59.3   &         45.5     &       40.1     &       42.8 
&        79.9   &         61.8     &       66.6     &       64.2 
\\ \bottomrule
\end{tabular}
\caption{\scriptsize F1 macro score results for SVM-linear classifier on multi-party framework fed by RE features on strongly (LOO CV) and weakly (3- and 1-shot) supervised scenarios. \textit{avg} column represents average values for weakly supervised scenario. Values with * represent when the results of REs with 1- or 3-shot training are higher than the best overall result of non RE methods for multi-party framework on strongly supervised scenario: DW with RF for EUS (\textit{70.1}) and GAL (\textit{71.3}); DW with LogReg for CAT (\textit{88.0}).}
\label{tab:f1_results_weakly}
\end{table*}

\section{Evaluation}

We evaluate the performance of the generated user representations with the aim of inferring political leaning among EUS, GAL and CAT regions in two distinct scenarios, which are defined on the amount of data used for training: strongly and weakly supervised scenarios.

\subsection{Strongly supervised scenario} We compare the performance of the diverse user representations combined with different classifiers in a strongly supervised scenario (leave-one-out cross-validation). Besides, results are also compared between binary (Table \ref{tab:f1_strong_bi}) and multi-party (Table \ref{tab:f1_strong_mul}) frameworks, empirically showing the challenges that involves shifting from binary to multi-class inference.

\paragraph{\textbf{Binary framework.}} Looking at the results reported in Table \ref{tab:f1_strong_bi}, we notice that each user representation model effectively captures political orientation through the left-right categorization, yielding high-performance results that depend on the employed classifier. However, it is evident that models trained using RE representations consistently demonstrate superior performance and stability in all regions, regardless of the classifier used.

\paragraph{\textbf{Multi-party framework.}} When analyzing the results presented in Table \ref{tab:f1_strong_mul}, it is evident that models trained using RE representations consistently demonstrate superior performance and stability across all regions. Among the models obtained with RE representations, the SVM classifiers obtains the best results. However, despite its popularity, the FA2 representations yield the lowest performance scores, indicating that they are the least suitable for this task. Both N2V and DW outperform FA2, but they are still surpassed by the models generated using RE representations.

As previously mentioned, FA2, DW and N2V can effectively capture information related to left-right orientation while they struggle when dealing with multi-class classifications. The notably superior results achieved within the binary framework compared to the multi-party framework illustrate that bipolar approaches to define political leaning lead to higher overall accuracy at the cost of essential nuances required to comprehend the specific political and social context. On the contrary, RE stands out as the most effective method for capturing finer-grained information related to more specific party-based political leanings, achieving high performance results in both the challenging multi-party context and the binary framework.

\subsection{Weakly supervised scenario} 

As a second step, the RE model is evaluated on the weakly supervised scenario to see the performance when considerably reducing training data. Table \ref{tab:f1_results_weakly} demonstrates that 2 dimensional representations obtained from REs through t-SNE and UMAP dimension reduction techniques yield superior results compared to representations generated by REs without any dimensionality reduction for both one-shot and three-shot settings. PCA may not be the most appropriate method for dimension reduction in this context, as it is unable to retain information specific to each community and yields inferior results compared to using the full-dimensional representations. In contrast, UMAP and t-SNE reductions outperform full-dimensional representations as they may preserve community related information due to their architecture based on nearest-neighbours.

Furthermore, the results obtained by combining REs with UMAP and t-SNE dimension reduction techniques in the weakly supervised scenario (Table \ref{tab:f1_results_weakly}) outperform any other model from the strongly supervised scenario (Table \ref{tab:f1_strong_mul}). Specifically, when REs are combined with UMAP, only one data point is necessary for training (one-shot) to outperform any other model of the strongly supervised scenario in more than 10 points at EUS and GAL and 7 points in CAT. We confirm that compressing RE representations into 2 dimensional features with UMAP or t-SNE is a good strategy to handle situations with very few annotated data, obtaining similar scores to those of RE on the strongly supervised scenario. Moreover, the results demonstrate consistency across all three regions, indicating that RE representations reach competitive and robust results even where only the user belonging to the political party is annotated.

\section{Discussion}

In this section, we will delve into a deeper analysis of the reported results by conducting an error analysis and generating visualizations of the user representations for the best method. 

\begin{figure}[h!]
\centering
    \includegraphics[width=0.33\linewidth]{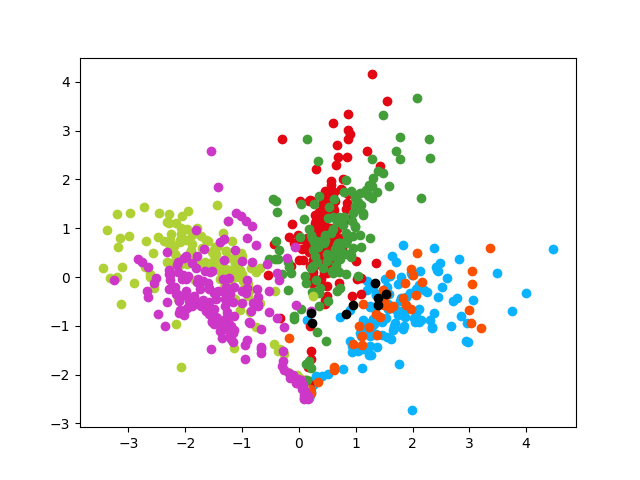} \hfil 
    \includegraphics[width=0.33\linewidth]{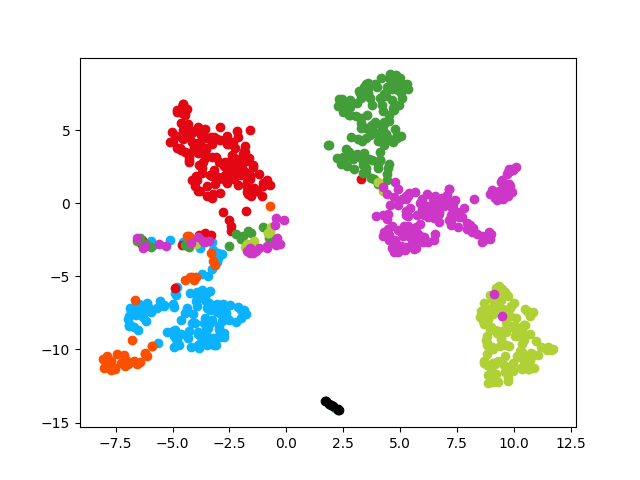}\hfil
    \includegraphics[width=0.32\linewidth]{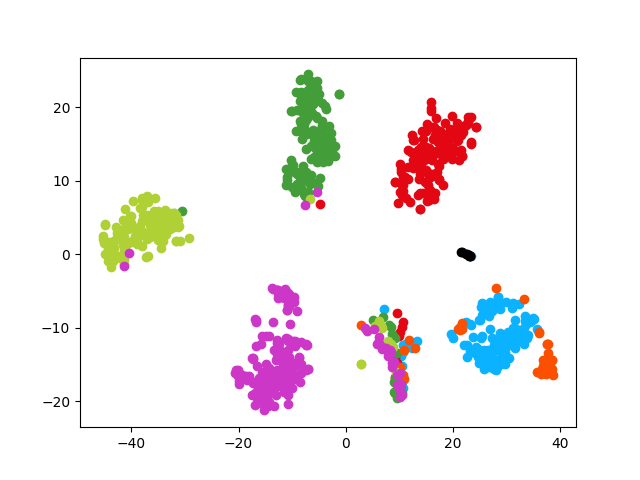}
    \caption{\scriptsize Visualization of PCA, UMAP and t-SNE 2 dimension reduction for EUS Relational Embedding representation.}
    \label{fig:emb_vis_eus} \hfil
\end{figure}

\begin{figure}[h!]
\centering
    \includegraphics[width=0.33\linewidth]{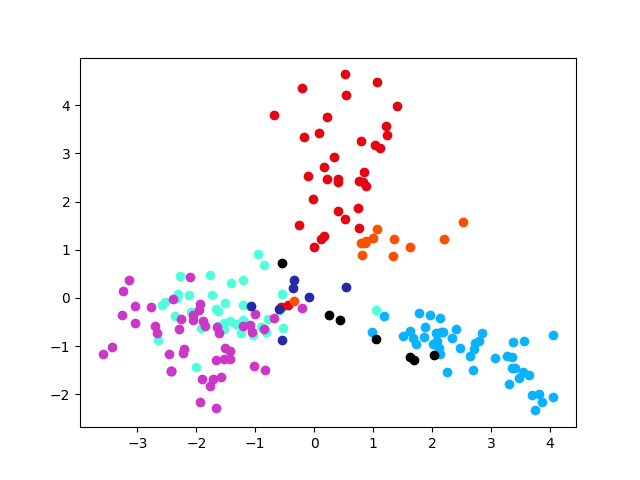}\hfil
    \includegraphics[width=0.33\linewidth]{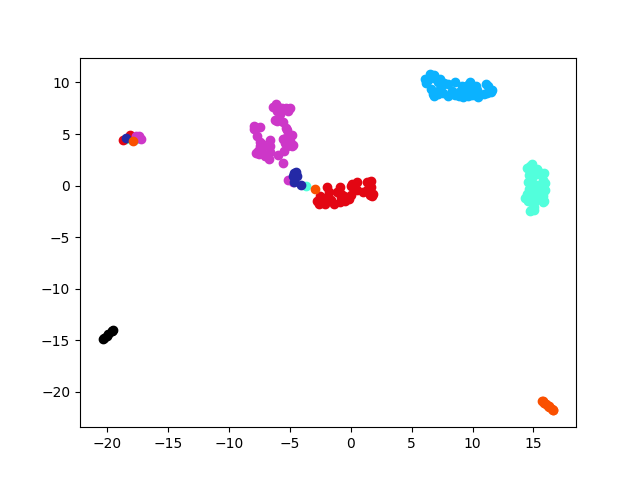}\hfil
    \includegraphics[width=0.33\linewidth]{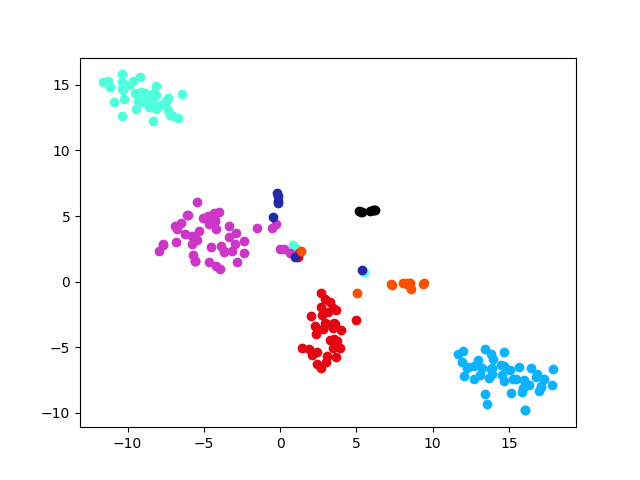}
    \caption{\scriptsize Visualization of PCA, UMAP and t-SNE 2 dimension reduction for GAL Relational Embedding representation.}
    \label{fig:emb_vis_gal} \hfil
\end{figure}

\begin{figure}[h!]
\centering
    \includegraphics[width=0.33\linewidth]{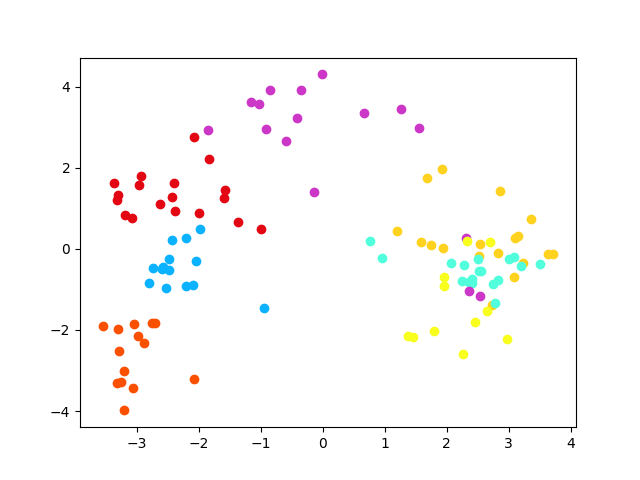}\hfil
    \includegraphics[width=0.33\linewidth]{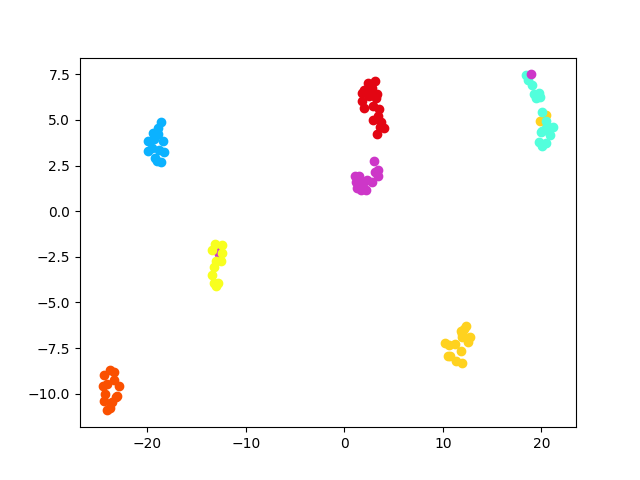}\hfil
    \includegraphics[width=0.33\linewidth]{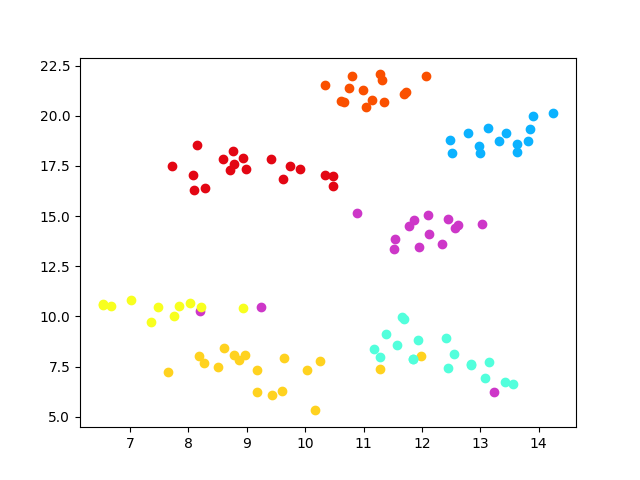}
    \caption{\scriptsize Visualization of PCA, UMAP and t-SNE 2 dimension reduction for CAT Relational Embedding representation.}
    \label{fig:emb_vis_cat} \hfil
\end{figure}

\paragraph{\textbf{Error analysis.}} In order to understand the considerable differences in performance among user representation methods, we conducted a detailed comparison of the best and second best methods. The confusion matrices presented in Figures \ref{fig:emb_vis_eus}, \ref{fig:emb_vis_gal} and \ref{fig:emb_vis_cat} report the errors performed by the Logistic Regression classifier using RE and DW user representation models for multi-party framework for the strongly supervised scenario.

\begin{figure}[h]
\centering
    \includegraphics[width=0.49\linewidth]{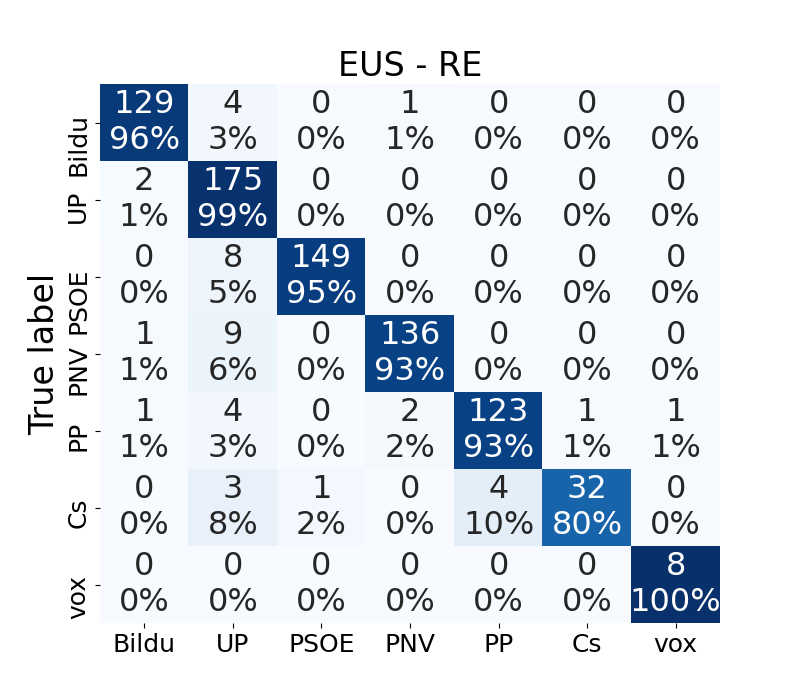} \hfil 
    \includegraphics[width=0.49\linewidth]{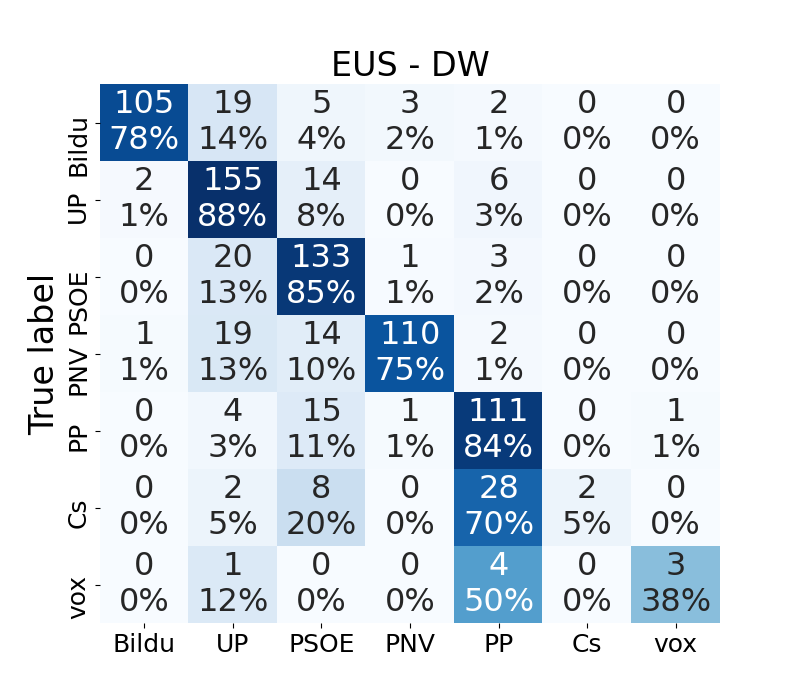}
    \caption{\scriptsize Confusion matrices for Logistic Regresion classifier using RE and DW user representation models in the strongly supervised scenario on the EUS dataset.}
    \label{fig:emb_cm_eus} \hfil
\end{figure}

In relation to the EUS dataset, the RE (Figure \ref{fig:emb_cm_eus} left) model shows classification errors that occur among classes such as Bildu-UP, PP-Cs and PSOE or PNV as UP. These errors take place among parties that are ideologically close, showing that the model fails between simmilar classes. On the other hand, the DW (Figure \ref{fig:emb_cm_eus} right) model makes a considerable number of errors across UP, PSOE and PNV, failing to classify users arround these orientations. Moreover, DW generally fails to classify right-wing unionist party members, grouping them all as PP users.

\begin{figure}[h]
\centering
    \includegraphics[width=0.49\linewidth]{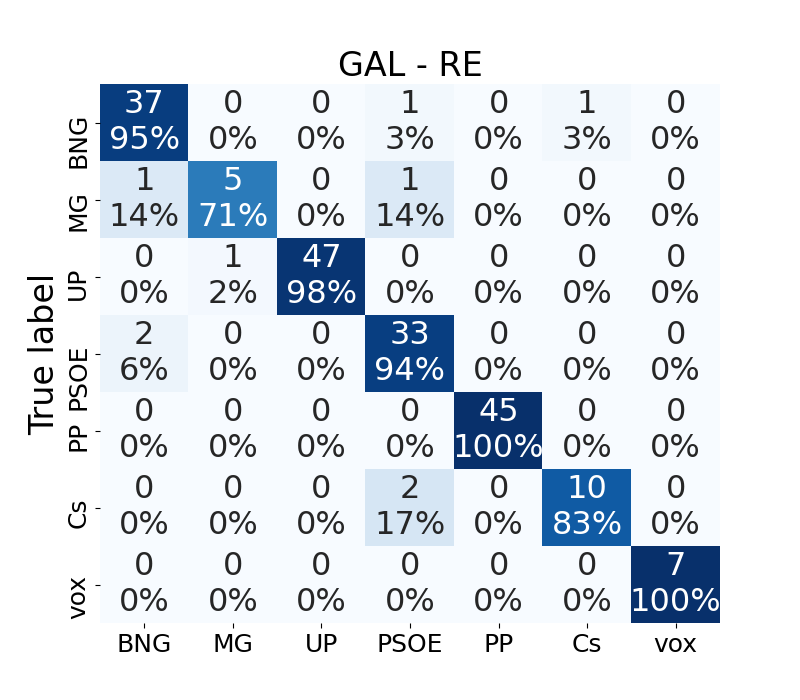} \hfil 
    \includegraphics[width=0.49\linewidth]{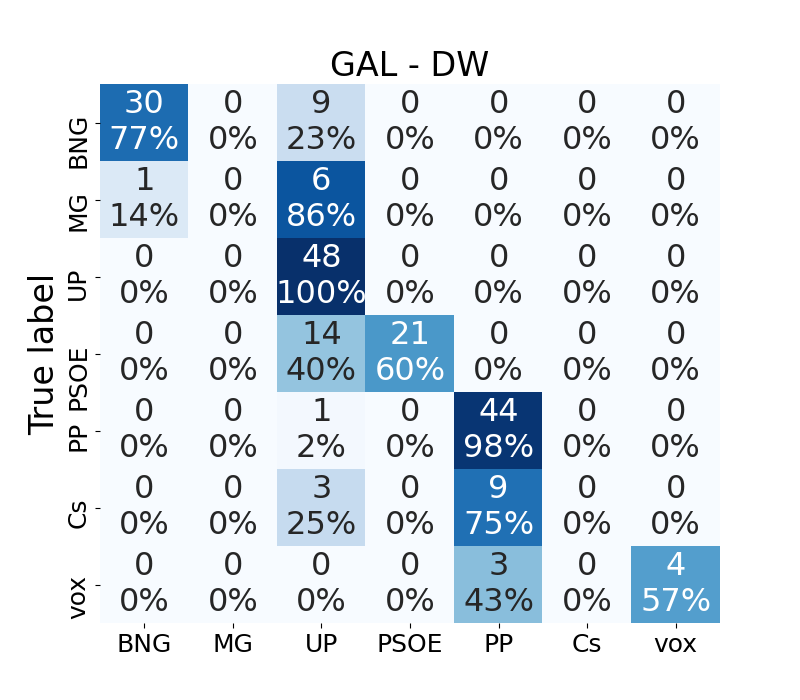}
    \caption{\scriptsize Confusion matrices for Logistic Regresion classifier using RE and DW user representation models in the strongly supervised scenario on the GAL dataset.}
    \label{fig:emb_cm_gal} \hfil
\end{figure}

When considering GAL dataset, DW (Figure \ref{fig:emb_cm_gal} right) misclassifies BNG, MG and PSOE users as UP members, grouping many left-wing representatives under the UP political orientation. Furthermore, some users from Vox and Cs are classified as PP users, being that party the main representative of the right-wing. The DW model seems to underrepresent political options in a left-right dichotomy, simplifying political complexity. On the other hand, the RE (Figure \ref{fig:emb_cm_gal} left) model has very few classification errors occurring among ideologically similar orientations, demonstrating the capacity of REs to represent parties as well as political orientations.

\begin{figure}[h]
\centering
    \includegraphics[width=0.49\linewidth]{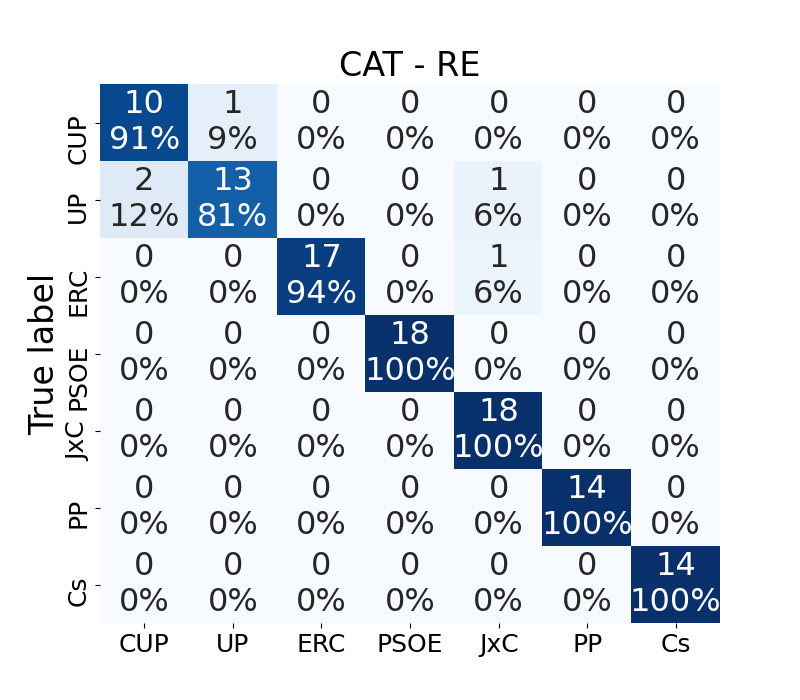} \hfil 
    \includegraphics[width=0.49\linewidth]{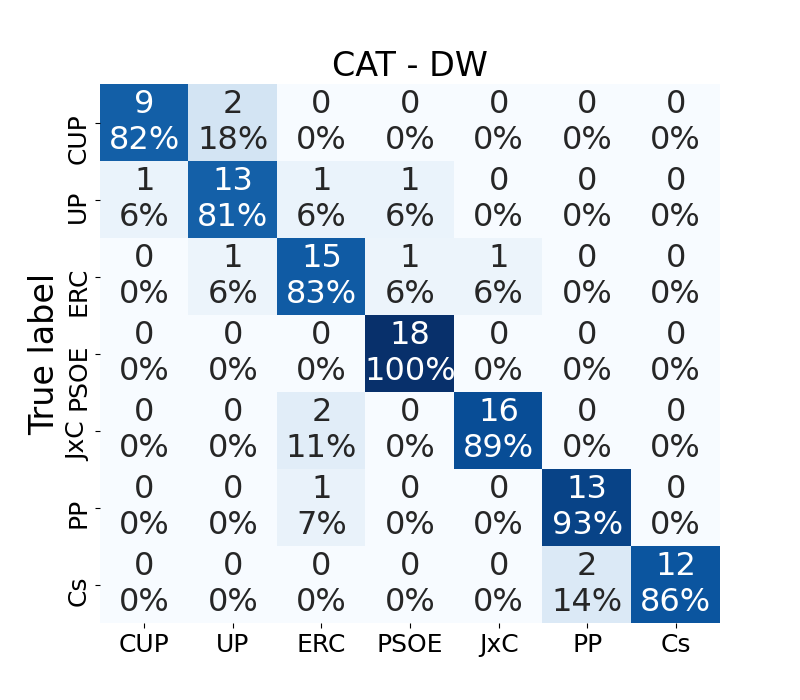}
    \caption{\scriptsize Confusion matrices for Logistic Regresion classifier using RE and DW user representation models in the strongly supervised scenario on the CAT dataset.}
    \label{fig:emb_cm_cat} \hfil
\end{figure}

Finally, with respect to CAT dataset, we can see that RE (Figure \ref{fig:emb_cm_cat} left) gets better results than DW (Figure \ref{fig:emb_cm_cat} right). However, both models achieve high performance despite minor deviations. This can be attributed to factors such as the smaller dataset size or balanced classes contributing to their outstanding performance.

Summarizing, classification errors usually take place among parties that are ideologically close, showing that retweet-based DW and RE models can capture general ideological tendencies. This conclusion aligns with the high results obtained by DW for the binary framework, confirming these models' ability to capture general ideological traits. However, the RE models show a high accuracy with a very low error rate, demonstrating that this model can also capture previously specified political parties besides the general ideological alignments. 

\paragraph{\textbf{Data visualization.}} In order to better understand the effectiveness of the RE user representation techniques, we visualize the RE representations of labeled users for the three regions, namely, EUS, GAL and CAT, by performing PCA, UMAP and t-SNE dimensionality reduction (as it was done in weakly supervised scenario). This visualization allows us to undertake a qualitative evaluation by correlating common-sense political knowledge to the representations obtained by our representation model.

\noindent\textit{\textbf{EUS} (Figure \ref{fig:emb_vis_eus}):} In the EUS dataset UMAP (Figure \ref{fig:emb_vis_eus} center) and t-SNE (Figure \ref{fig:emb_vis_eus} right) visualizations have similar groupings with clear clusters for specific political organizations, which are arranged based on their political proximity. Thus, the centered positions of the graph are taken by the parties in the Basque government, formed by PNV (\textcolor{pnv}{$\bullet$}) and PSOE (\textcolor{psoe}{$\bullet$}). The leftist positions are represented by UP (\textcolor{up}{$\bullet$}) and Bildu (\textcolor{bildu}{$\bullet$}), located on one side. Whereas right-winged positions represented by PP (\textcolor{pp}{$\bullet$}) and Cs (\textcolor{cs}{$\bullet$}), are pictured on the other side, while alt-right Vox ({$\bullet$}) is represented as an outlier. PCA visualization (Figure \ref{fig:emb_vis_eus} left) is fuzzier but able to group users on 3 different groups corresponding to the previous global views; PNV and PSOE representing centered positions; Bildu and UP representing the left and progressives;  PP, Cs and Vox representing the right and the conservatives.

\noindent\textit{\textbf{GAL} (Figure \ref{fig:emb_vis_gal}):} In this dataset t-SNE visualizations (Figure \ref{fig:emb_vis_gal} right) show clear clusters for the political parties, drawing them on a singular axis. On one end we have BNG (\textcolor{bng}{$\bullet$}) representing a pro-independence left-wing followed by UP (\textcolor{up}{$\bullet$}) and MG (\textcolor{mg}{$\bullet$}) representing the left; PSOE (\textcolor{psoe}{$\bullet$}) and Cs (\textcolor{cs}{$\bullet$}) represent centered positions next to PP (\textcolor{pp}{$\bullet$}) representing the right and the conservatives on the oposite end. PCA visualization (Figure \ref{fig:emb_vis_gal} left) also groups the users on 3 different groups mixing political parties into a more simple layout; UP, BNG and MG as left and progressives; PSOE and Cs grouped on centered positions; PP (\textcolor{pp}{$\bullet$}) and Vox ($\bullet$) grouped as right and the conservatives.

\noindent\textit{\textbf{CAT} (Figure \ref{fig:emb_vis_cat}):} 
In the CAT dataset PCA (Figure \ref{fig:emb_vis_cat} left) and t-SNE (Figure \ref{fig:emb_vis_cat} right) visualizations provide symbolic representations of the political reality, wherein political parties are positioned close to each other based on their stance regarding the independence process. On one side, parties like Cs (\textcolor{cs}{$\bullet$}) and PP (\textcolor{pp}{$\bullet$}), which strongly advocate for unity with the Spanish kingdom, are clustered in the left (PCA) or top (t-SNE) portions of the plot. Conversely, pro-independence parties such as CUP (\textcolor{cup}{$\bullet$}), JxC (\textcolor{jxc}{$\bullet$}), and ERC (\textcolor{erc}{$\bullet$}) are clustered on the right (PCA) or bottom (t-SNE) side. In more central positions, UP (\textcolor{up}{$\bullet$}) and PSOE (\textcolor{psoe}{$\bullet$}) are situated between both groupings, serving as a connecting links bridging the divide between the two sides.

It seems that the dataset size has an impact on the visualizations, with representations becoming fuzzier as the dataset size increases. Being the biggest dataset, EUS exhibits the largest degree of fuzziness, while CAT is the smallest dataset and has the most defined communities. Additionally, the clarity of the communities depending on the dimensionality reduction technique is in accordance with the results achieved by these dimension reduction techniques on the weakly supervised scenario. On the one hand, the PCA dimension reduction technique does not clearly discriminate the communities related to the specific political parties. However, it can effectively display information related to more general political orientations by clustering ideologically similar parties closely together in the same euclidean space. On the other hand, in the graphs arisen from UMAP and t-SNE dimension reduction techniques, it can be seen that the communities are clearly defined and situated depending on their ideological similarities. Regardless of the dimension reduction technique or the dataset size, REs can effectively represent the political communities as well as the ideological similarities and disparities among them. These findings suggest that RE user representations have the capacity to embed knowledge about the socio-political environment leaning on retweet based user interactions.

\section{Conclusion and Future Work}

In this work we have explored the ability to dynamically infer the political leaning of social media users across left-right and multi party-based frameworks which, to the best of our knowledge, has not been studied before. In order to compare binary and multi-party frameworks, we compile a dataset labeling users according to both their affiliation with a political party and their left-right orientation. We collect data from three politically complex areas in Spain, characterized by plurinationality and the emergence of new political actors. To conduct dynamic political leaning inference we propose a two-step approach, starting with different unsupervised user representations through retweets, followed by political party classification. When assessing the performances of user representation methods among different frameworks, it is observed that while performances are high at the binary framework, there is a significant decrease for baseline methods in the multi-party framework. We believe that this illustrates the challenges associated with inferring political leanings in multi-party frameworks.

Nonetheless, results indicate that Relational Embeddings, when combined with any classifier, yield high-performance results across all regions and evaluation scenarios. Thus, even when only a single user belonging to each political party is annotated, REs in combination with SVM-linear classifier outperforms any other baseline model from the strongly supervised multi-party framework. Furthermore, error analysis and visual representations reveal that REs can effectively capture political affinities within and across political leanings. These considerations underscore the adaptability of REs and its potential to capture socio-political information in extremely diverse and dynamic real-world situations.

Considering this, we have a keen interest in applying interaction-based user representation techniques to other tasks such as propaganda or misinformation detection. Additionally, we want to include additional data sources for user representation, such as textual data, in order to extend the user representation idea to social media platforms where retweet interactions are not available.



\begin{thebibliography}{49}

\bibitem[\protect\citename{Agerri \bgroup et al.\egroup }2021]{vaxxstance2021}
Agerri, R., R.~Centeno, M.~Espinosa, J.~Fernandez~de Landa, and A.~Rodrigo.
\newblock 2021.
\newblock {VaxxStance@IberLEF 2021: Overview of the Task on Going Beyond Text in Cross-Lingual Stance Detection}.
\newblock {\em Procesamiento del Lenguaje Natural}, 67:173--181.

\bibitem[\protect\citename{Alkhalifa and Zubiaga}2020]{QMUL-SDS}
Alkhalifa, R. and A.~Zubiaga.
\newblock 2020.
\newblock {QMUL-SDS@SardiStance: Leveraging Network Inter-actions to Boost Performance on Stance Detection using Knowledge Graphs.}
\newblock In {\em {Proceedings of the 7th Evaluation Campaign of Natural Language Processing and Speech Tools for Italian (EVALITA 2020)}}. {CEUR} Workshop Proceedings.

\bibitem[\protect\citename{Barber{\'a}}2015]{Barber2015BirdsOT}
Barber{\'a}, P.
\newblock 2015.
\newblock Birds of the same feather tweet together: Bayesian ideal point estimation using twitter data.
\newblock {\em Political Analysis}, 23:76 -- 91.

\bibitem[\protect\citename{Barber{\'a} \bgroup et al.\egroup }2015]{barbera2015tweeting}
Barber{\'a}, P., J.~T. Jost, J.~Nagler, J.~A. Tucker, and R.~Bonneau.
\newblock 2015.
\newblock Tweeting from left to right: Is online political communication more than an echo chamber?
\newblock {\em Psychological science}, 26(10):1531--1542.

\bibitem[\protect\citename{Barber{\'a} and Rivero}2015]{Barber2015UnderstandingTP}
Barber{\'a}, P. and G.~Rivero.
\newblock 2015.
\newblock Understanding the political representativeness of twitter users.
\newblock {\em Social Science Computer Review}, 33:712 -- 729.

\bibitem[\protect\citename{Boutet, Kim, and Yoneki}2012]{Boutet2012WhatsIY}
Boutet, A., H.~Kim, and E.~Yoneki.
\newblock 2012.
\newblock What’s in your tweets? i know who you supported in the uk 2010 general election.
\newblock {\em Proceedings of the International AAAI Conference on Web and Social Media}, 6:411--414.

\bibitem[\protect\citename{Cignarella \bgroup et al.\egroup }2020]{cignarella2020sardistance}
Cignarella, A.~T., M.~Lai, C.~Bosco, V.~Patti, and P.~Rosso.
\newblock 2020.
\newblock {SardiStance@EVALITA2020: Overview of the Task on Stance Detection in Italian Tweets}.
\newblock In V.~Basile, D.~Croce, M.~Di~Maro, and L.~C. Passaro, editors, {\em {Proceedings of the 7th Evaluation Campaign of Natural Language Processing and Speech Tools for Italian (EVALITA 2020)}}. CEUR-WS.org.

\bibitem[\protect\citename{CIS}2019]{cis_es}
CIS.
\newblock 2019.
\newblock {\em Barómetro de diciembre 2019. Postelectoral Elecciones Generales 2019}, volume 3269.
\newblock Centro de Investigaciones Sociológicas, 11.

\bibitem[\protect\citename{CIS}2020a]{cis_gal}
CIS.
\newblock 2020a.
\newblock {\em Preelectoral de Galicia. Elecciones Autonómicas julio 2020}, volume 3287.
\newblock Centro de Investigaciones Sociológicas, 06.

\bibitem[\protect\citename{CIS}2020b]{cis_eus}
CIS.
\newblock 2020b.
\newblock {\em Preelectoral del País Vasco. Elecciones Autonómicas julio 2020}, volume 3286.
\newblock Centro de Investigaciones Sociológicas, 06.

\bibitem[\protect\citename{Conover \bgroup et al.\egroup }2011a]{conover2011predicting}
Conover, M.~D., B.~Gon{\c{c}}alves, J.~Ratkiewicz, A.~Flammini, and F.~Menczer.
\newblock 2011a.
\newblock Predicting the political alignment of twitter users.
\newblock In {\em 2011 IEEE third international conference on privacy, security, risk and trust and 2011 IEEE third international conference on social computing}, pages 192--199. IEEE.

\bibitem[\protect\citename{Conover \bgroup et al.\egroup }2011b]{conover2011political}
Conover, M.~D., J.~Ratkiewicz, M.~Francisco, B.~Gon{\c{c}}alves, F.~Menczer, and A.~Flammini.
\newblock 2011b.
\newblock Political {P}olarization on {T}witter.
\newblock In {\em Proceedings of the International AAAI Conference on Web and Social Media}.

\bibitem[\protect\citename{Darwish \bgroup et al.\egroup }2020]{darwish2020unsupervised}
Darwish, K., P.~Stefanov, M.~Aupetit, and P.~Nakov.
\newblock 2020.
\newblock Unsupervised user stance detection on twitter.
\newblock In {\em Proceedings of the International AAAI Conference on Web and Social Media}, volume~14, pages 141--152.

\bibitem[\protect\citename{Fernandez~de Landa and Agerri}2021]{basqueyoungtwitter}
Fernandez~de Landa, J. and R.~Agerri.
\newblock 2021.
\newblock Social analysis of young basque-speaking communities in twitter.
\newblock {\em Journal of Multilingual and Multicultural Development}, 0(0):1--15.

\bibitem[\protect\citename{Fernandez~de Landa and Agerri}2022]{fernandez2022relational}
Fernandez~de Landa, J. and R.~Agerri.
\newblock 2022.
\newblock Relational embeddings for language independent stance detection.
\newblock {\em arXiv e-prints}, pages arXiv--2210.

\bibitem[\protect\citename{Fernandez~de Landa, Agerri, and Alegria}2019]{fernandez2019large}
Fernandez~de Landa, J., R.~Agerri, and I.~Alegria.
\newblock 2019.
\newblock Large {S}cale {L}inguistic {P}rocessing of {T}weets to {U}nderstand {S}ocial {I}nteractions among {S}peakers of {L}ess {R}esourced {L}anguages: {T}he {B}asque {C}ase.
\newblock {\em Information}, 10(6):212.

\bibitem[\protect\citename{Fernandez~de Landa, Zubiaga, and Agerri}2023]{FdLPolLeaIn}
Fernandez~de Landa, J., A.~Zubiaga, and R.~Agerri.
\newblock 2023.
\newblock Generalizing political leaning inference to multi-party systems: Insights from the uk political landscape.
\newblock {\em ArXiv}, abs/2312.01738.

\bibitem[\protect\citename{Ford and Jennings}2020]{ford2020changing}
Ford, R. and W.~Jennings.
\newblock 2020.
\newblock The changing cleavage politics of western europe.
\newblock {\em Annual review of political science}, 23:295--314.

\bibitem[\protect\citename{Fruchterman and Reingold}1991]{Fruchterman1991GraphDB}
Fruchterman, T. M.~J. and E.~M. Reingold.
\newblock 1991.
\newblock Graph drawing by force‐directed placement.
\newblock {\em Software: Practice and Experience}, 21.

\bibitem[\protect\citename{Garimella and Weber}2017]{garimella2017long}
Garimella, V. R.~K. and I.~Weber.
\newblock 2017.
\newblock A long-term analysis of polarization on twitter.
\newblock In {\em Proceedings of the International AAAI Conference on Web and Social Media}, volume~11.

\bibitem[\protect\citename{Grover and Leskovec}2016]{Grover2016node2vecSF}
Grover, A. and J.~Leskovec.
\newblock 2016.
\newblock node2vec: Scalable feature learning for networks.
\newblock {\em Proceedings of the 22nd ACM SIGKDD International Conference on Knowledge Discovery and Data Mining}.

\bibitem[\protect\citename{Hua, Ristenpart, and Naaman}2020]{Hua2020TowardsMA}
Hua, Y., T.~Ristenpart, and M.~Naaman.
\newblock 2020.
\newblock Towards measuring adversarial twitter interactions against candidates in the us midterm elections.
\newblock In {\em ICWSM}.

\bibitem[\protect\citename{Jacomy \bgroup et al.\egroup }2014]{jacomy2014forceatlas2}
Jacomy, M., T.~Venturini, S.~Heymann, and M.~Bastian.
\newblock 2014.
\newblock Force{A}tlas2, a continuous graph layout algorithm for handy network visualization designed for the {G}ephi software.
\newblock {\em PloS one}, 9(6):e98679.

\bibitem[\protect\citename{Jusup \bgroup et al.\egroup }2022]{socialphysics2022}
Jusup, M., P.~Holme, K.~Kanazawa, M.~Takayasu, I.~Romić, Z.~Wang, S.~Geček, T.~Lipić, B.~Podobnik, L.~Wang, W.~Luo, T.~Klanjšček, J.~Fan, S.~Boccaletti, and M.~Perc.
\newblock 2022.
\newblock Social physics.
\newblock {\em Physics Reports}, 948:1--148.
\newblock Social physics.

\bibitem[\protect\citename{Keating}2001]{KeatingMichael}
Keating, M.
\newblock 2001.
\newblock {\em {Plurinational Democracy: Stateless Nations in a Post-Sovereignty Era}}.
\newblock Oxford University Press, 11.

\bibitem[\protect\citename{Kermani and Adham}2021]{kermani2021mapping}
Kermani, H. and M.~Adham.
\newblock 2021.
\newblock Mapping persian twitter: Networks and mechanism of political communication in iranian 2017 presidential election.
\newblock {\em Big Data \& Society}, 8(1):20539517211025568.

\bibitem[\protect\citename{Kotroyannos, Tzagkarakis, and Pappas}2018]{kotroyannos2018south}
Kotroyannos, D., S.~I. Tzagkarakis, and I.~Pappas.
\newblock 2018.
\newblock South european populism as a consequence of the multidimensional crisis? the cases of syriza, podemos and m5s.
\newblock {\em European Quarterly of Political Attitudes and Mentalities}, 7(4):1--18.

\bibitem[\protect\citename{Leong \bgroup et al.\egroup }2020]{Leong2020ConservativeAL}
Leong, Y.~C., J.~Chen, R.~Willer, and J.~Zaki.
\newblock 2020.
\newblock Conservative and liberal attitudes drive polarized neural responses to political content.
\newblock {\em Proceedings of the National Academy of Sciences}, 117:27731 -- 27739.

\bibitem[\protect\citename{Lisi}2018]{lisi2018party}
Lisi, M.
\newblock 2018.
\newblock {\em Party system change, the European crisis and the state of democracy}.
\newblock Routledge.

\bibitem[\protect\citename{Ma \bgroup et al.\egroup }2021]{9565320}
Ma, X., J.~Wu, S.~Xue, J.~Yang, C.~Zhou, Q.~Z. Sheng, H.~Xiong, and L.~Akoglu.
\newblock 2021.
\newblock A comprehensive survey on graph anomaly detection with deep learning.
\newblock {\em IEEE Transactions on Knowledge and Data Engineering}, pages 1--1.

\bibitem[\protect\citename{Magdy \bgroup et al.\egroup }2016]{magdy16}
Magdy, W., K.~Darwish, N.~Abokhodair, A.~Rahimi, and T.~Baldwin.
\newblock 2016.
\newblock \#isisisnotislam or \#deportallmuslims? predicting unspoken views.
\newblock In {\em Proceedings of the 8th ACM Conference on Web Science}, WebSci '16, page 95–106, New York, NY, USA. Association for Computing Machinery.

\bibitem[\protect\citename{Makazhanov and Rafiei}2013]{Makazhanov2013PredictingPP}
Makazhanov, A. and D.~Rafiei.
\newblock 2013.
\newblock Predicting political preference of twitter users.
\newblock {\em Social Network Analysis and Mining}, 4:1--15.

\bibitem[\protect\citename{McGann, Dellepiane-Avellaneda, and Bartle}2019]{MCGANN201948}
McGann, A., S.~Dellepiane-Avellaneda, and J.~Bartle.
\newblock 2019.
\newblock Parallel lines? policy mood in a plurinational democracy.
\newblock {\em Electoral Studies}, 58:48--57.

\bibitem[\protect\citename{McInnes \bgroup et al.\egroup }2018]{McInnes2018UMAPUM}
McInnes, L., J.~Healy, N.~Saul, and L.~Gro{\ss}berger.
\newblock 2018.
\newblock Umap: Uniform manifold approximation and projection.
\newblock {\em J. Open Source Softw.}, 3:861.

\bibitem[\protect\citename{Mikolov \bgroup et al.\egroup }2013]{mikolov2013efficient}
Mikolov, T., K.~Chen, G.~Corrado, and J.~Dean.
\newblock 2013.
\newblock Efficient estimation of word representations in vector space.
\newblock {\em arXiv preprint arXiv:1301.3781}.

\bibitem[\protect\citename{Morales, Monti, and Starnini}2021]{Morales2021NoEI}
Morales, G. D.~F., C.~Monti, and M.~Starnini.
\newblock 2021.
\newblock No echo in the chambers of political interactions on reddit.
\newblock {\em Scientific Reports}, 11.

\bibitem[\protect\citename{Morlino and Raniolo}2017]{morlino2017impact}
Morlino, L. and F.~Raniolo.
\newblock 2017.
\newblock {\em The impact of the economic crisis on South European democracies}.
\newblock Springer.

\bibitem[\protect\citename{Pedregosa \bgroup et al.\egroup }2011]{scikit-learn}
Pedregosa, F., G.~Varoquaux, A.~Gramfort, V.~Michel, B.~Thirion, O.~Grisel, M.~Blondel, P.~Prettenhofer, R.~Weiss, V.~Dubourg, J.~Vanderplas, A.~Passos, D.~Cournapeau, M.~Brucher, M.~Perrot, and E.~Duchesnay.
\newblock 2011.
\newblock Scikit-learn: Machine learning in python.
\newblock {\em Journal of Machine Learning Research}, 12:2825--2830.

\bibitem[\protect\citename{Pennacchiotti and Popescu}2011]{Pennacchiotti2011DemocratsRA}
Pennacchiotti, M. and A.~M. Popescu.
\newblock 2011.
\newblock Democrats, republicans and starbucks afficionados: user classification in twitter.
\newblock In {\em KDD}.

\bibitem[\protect\citename{Perozzi, Al-Rfou, and Skiena}2014]{deepwalk}
Perozzi, B., R.~Al-Rfou, and S.~Skiena.
\newblock 2014.
\newblock Deepwalk: Online learning of social representations.
\newblock KDD '14, page 701–710. Association for Computing Machinery.

\bibitem[\protect\citename{Preotiuc-Pietro \bgroup et al.\egroup }2017]{PreotiucPietro2017BeyondBL}
Preotiuc-Pietro, D., Y.~Liu, D.~J. Hopkins, and L.~H. Ungar.
\newblock 2017.
\newblock Beyond binary labels: Political ideology prediction of twitter users.
\newblock In {\em ACL}.

\bibitem[\protect\citename{Rama, Cordero, and Zagórski}2021]{endbipartspain}
Rama, J., G.~Cordero, and P.~Zagórski.
\newblock 2021.
\newblock Three is a crowd? podemos, ciudadanos, and vox: The end of bipartisanship in spain.
\newblock {\em Frontiers in Political Science}, 3.

\bibitem[\protect\citename{Rashed \bgroup et al.\egroup }2021]{Rashed2021EmbeddingsBasedCF}
Rashed, A., M.~Kutlu, K.~Darwish, T.~Elsayed, and C.~Bayrak.
\newblock 2021.
\newblock Embeddings-based clustering for target specific stances: The case of a polarized turkey.
\newblock In {\em ICWSM}.

\bibitem[\protect\citename{Recuero, Zago, and Soares}2019]{recuero2019using}
Recuero, R., G.~Zago, and F.~Soares.
\newblock 2019.
\newblock Using social network analysis and social capital to identify user roles on polarized political conversations on twitter.
\newblock {\em Social Media+ Society}, 5(2):2056305119848745.

\bibitem[\protect\citename{Soares and Recuero}2021]{soares2021hashtag}
Soares, F.~B. and R.~Recuero.
\newblock 2021.
\newblock Hashtag wars: Political disinformation and discursive struggles on twitter conversations during the 2018 brazilian presidential campaign.
\newblock {\em Social Media+ Society}, 7(2):20563051211009073.

\bibitem[\protect\citename{Stefanov \bgroup et al.\egroup }2020]{Stefanov2020PredictingTT}
Stefanov, P., K.~Darwish, A.~Atanasov, and P.~Nakov.
\newblock 2020.
\newblock Predicting the topical stance and political leaning of media using tweets.
\newblock In {\em ACL}.

\bibitem[\protect\citename{Van~der Maaten and Hinton}2008]{van2008visualizing}
Van~der Maaten, L. and G.~Hinton.
\newblock 2008.
\newblock Visualizing data using t-sne.
\newblock {\em Journal of machine learning research}, 9(11).

\bibitem[\protect\citename{Xiao \bgroup et al.\egroup }2020]{timme20}
Xiao, Z., W.~Song, H.~Xu, Z.~Ren, and Y.~Sun.
\newblock 2020.
\newblock Timme: Twitter ideology-detection via multi-task multi-relational embedding.
\newblock In {\em Proceedings of the 26th ACM SIGKDD International Conference on Knowledge Discovery \& Data Mining}, KDD '20, page 2258–2268, New York, NY, USA. Association for Computing Machinery.

\bibitem[\protect\citename{Zubiaga \bgroup et al.\egroup }2019]{Zubiaga2019PoliticalHI}
Zubiaga, A., B.~Wang, M.~Liakata, and R.~Procter.
\newblock 2019.
\newblock Political homophily in independence movements: Analyzing and classifying social media users by national identity.
\newblock {\em IEEE Intelligent Systems}, 34:34--42.

\end{thebibliography}

\appendix

\end{document}